\documentclass[floatfix,showpacs,twocolumn,aps,amsmath,amssymb,nofootinbib]{revtex4}
\usepackage{graphicx}
\usepackage[tight]{subfigure}

\newcommand{\abs}[1]{\lvert#1\rvert}

\newcommand{\E}{\mathcal{E}}
\renewcommand{\L}{\mathcal{L}}
\newcommand{\W}{\mathcal{W}} \newcommand{\ang}[1]{\overset{\circ}{#1}}
\newcommand{\Eo}{\ang{\E}} \newcommand{\Lo}{\ang{\L}}
\newcommand{\Ero}{\ang{E}} \newcommand{\Lro}{\ang{L}}

\newcommand{\F}[1]{\mathcal{F}\left[#1\right]}

\newcommand{\figwidth}{2.8in}

\begin{document}
\title{Stability of Negative Image Equilibria in Spike-Timing
  Dependent Plasticity} \author{Alan Williams}
\email{williaal@ohsu.edu} \author{Patrick D. Roberts}
\email{robertpa@ohsu.edu} \affiliation{Neurological Sciences
  Institute, Oregon Health \& Science University, 505 NW 185th Avenue, Beaverton, OR 97006} \author{Todd
  K. Leen} \email{tleen@cse.ogi.edu} \affiliation{Department of
  Computer Science and Engineering, OGI School of Science \&
  Engineering, Oregon Health \& Science University}
 
Classification Scheme.

\date{\today}

\begin{abstract}
We investigate the stability of negative image equilibria in mean
synaptic weight dynamics governed by spike-timing dependent plasticity (STDP). The neural architecture of the model is based on the
electrosensory lateral line lobe (ELL) of mormyrid electric fish,
which forms a negative image of the reafferent signal from the fish's
own electric discharge to optimize detection of external electric
fields. We derive a necessary and sufficient condition for stability,
for arbitrary postsynaptic potential functions and arbitrary learning
rules. We then apply the general result to several examples of
biological interest.

\end{abstract}

\maketitle

\section{Introduction}
Synaptic plasticity is thought to be a fundamental mechanism for
learning and adaptation in biological neural networks
\cite{Hebb49}. The activity dependence of synaptic plasticity has been
observed experimentally \cite{Lomo71,Bliss73}, but the precise nature
of that dependence, and its functional or computational consequences,
are still largely unknown. The purpose of the present article is to
derive clear functional consequences from specific forms of
activity-dependent synaptic plasticity.

Current models of synaptic plasticity are of two main types:
rate-based, and timing-based. In rate-based models, changes in
synaptic weight depend on the mean spike rate of presynaptic and
postsynaptic cells, usually via correlations
\cite{Sejnowski77,Bienenstock82}. Since mean spike rates are averages
over time windows containing many spikes, the timing of
individual spikes is unimportant in rate-based models. Recent experimental studies \cite{Markram97a,Bell97a,Bi98} have shown that in some systems the precise timing of individual spikes can have a pronounced effect on synaptic
plasticity. Models of such \emph{spike-timing dependent plasticity}
(STDP) \cite{Abbott00} calculate changes in synaptic weights by
combining the effect of all pairs of presynaptic and postsynaptic
spikes \cite{vanRossum00,Rubin01,Yoshioka02,Zhigulin03,Cateau03}, where the effect of each pair is a function of the time between
them (called the spike-timing dependent learning rule).

One system in which STDP has been observed experimentally, and where
its functional role is understood, is the electrosensory lateral line
lobe (ELL) of mormyrid electric fish \cite{Bell97a}. The mormyrid
identifies objects in its environment by emitting a stereotyped
electrical discharge and detecting the perturbations to the resulting
electrical field at the skin surface due to external objects. To
cancel the predictable sensory input due to its own discharge, the
mormyrid sends a series of time-delayed, time-locked inputs to the
ELL, synchronized to the fish's electrical discharge \cite{Bell92a}. The neurons
receiving these inputs have plastic synapses onto neurons receiving 
primary afferent input. The repeated time-locked inputs, paired with
the reafferent input, act via a spike-timing dependent learning rule
to change synaptic weights, in such a way that the summed postsynaptic
potential due to the time-locked inputs forms a \emph{negative image}
of the potential due to the fish's own discharge \cite{Bell97b}. This
effectively nulls out the sensory effect of the fish's own discharge, thus improving detectability
of perturbations due to external objects.

To be behaviorally useful to the fish, the set of synaptic weights
which create the negative image must be a stable equilibrium for the
weight dynamics induced by the spike-timing dependent learning
rule. Conditions for existence and stability of such equilibria were
first explored in \cite{Roberts00b}; the present paper is an extension
and refinement of that work. The principal extension is the derivation
of an analytic criterion for stability of negative image equilibria
for arbitrary postsynaptic potential functions and arbitrary
spike-timing dependent learning rules.
 
\section{Framework}
\begin{figure}
\begin{center}
\includegraphics[width=1.5in]{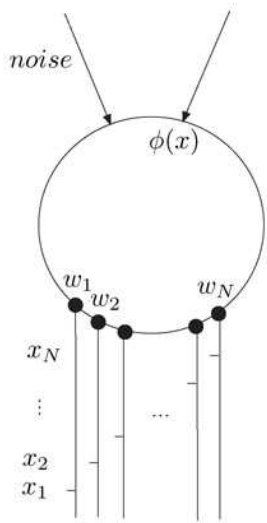}
\caption{Schematic of the architecture. The postsynaptic cell receives
  inputs from $N$ presynaptic neurons, a repeated external input
  $\phi(x)$, and unspecified noisy inputs. Presynaptic neuron $i$
  spikes at time $x_i$ in each period of $\phi$, and has synaptic
  weight $w_i$ onto the postsynaptic cell. }
\label{figschematic}
\end{center}
\end{figure}

The model consists of a single postsynaptic cell driven by an array of
time-locked presynaptic cells, a repeated external input, and other
unspecified inputs collectively modeled as a single noisy external
input \cite{Kempter99a, Roberts99a, Roberts00a} (Fig. \ref{figschematic}). This architecture is based on the mormyrid
ELL, but is general enough to capture the dynamics of other neural
systems hypothesized to have an array of time-delayed, time-locked
inputs \cite{Hahnloser02,Ehrlich97}.
The framework for the neural dynamics is the spike response model
\cite{Gerstner93}, without refractoriness.

Each presynaptic cell $i$ spikes exactly once at a fixed time within
each sweep of the repeated external input, causing a corresponding
postsynaptic potential response (PSP) in the postsynaptic cell.

The total membrane potential in the postsynaptic cell is the sum of
these PSPs, weighted by synaptic efficacies (weights) $w_i$, and the
two external inputs. This membrane potential induces the postsynaptic
cell to spike at a certain (noisy) rate. Each presynaptic spike causes
a constant (nonassociative) change in the weight $w_i$, and each
postsynaptic and presynaptic spike pair causes a change in $w_i$
according to a spike-timing dependent learning rule, namely a function
of the time difference between the postsynaptic and presynaptic spikes
(associative learning).

The repeated external input is modeled as a periodic input with period
T. We use two time variables: $x\in [0,T)$ for the time within each
  repetition of the external input, and $t=nT$, $n\in \mathbb{Z}$ for
  the time of initiation of each such period \cite{Roberts99a,
    Roberts00a, Roberts00c}. General dynamical quantities will be
  functions of the pair $(x,t)$. Let $x_i$ be the time within each
  period when presynaptic cell $i$ spikes, and $w_i(x,t)$ its
  corresponding weight. Since presynaptic spikes are time-locked to
  the external input, $x_i$ is independent of $t$. Let $\E(s)$ be the
  PSP evoked by neuron $i$ at time $s$ after a spike. We assume $\E$
  is causal: $\E(s)=0$ for $s<0$. Let $\alpha$ be the nonassociative
  weight change due to a presynaptic spike, and $\L(s)$ the
  associative weight change due to a postsynaptic spike time $s$ after
  a presynaptic spike. Let $\phi(x)$ be the periodic external input,
  and $U(x,t)$ the total postsynaptic potential due to the non-noisy
  inputs. We assume that for each $t$, the mean instantaneous
  postsynaptic spike rate density (in $x$) is given by $f(U(x,t))$ for
  some positive and strictly increasing function $f$. The function $f$
  can be thought of as the effective gain of the postsynaptic cell in
  the presence of the noisy inputs. High or low noise correspond to an
  $f$ with small or large maximum slope respectively. No attempt is
  made to include a refractory period for postsynaptic spikes; and we
  will assume the period of $\phi$ is greater than the refractory
  period of the presynaptic neurons, so that refractoriness on the
  presynaptic side is irrelevant.
                                                                                                          
Changes in weights will be implemented as discrete steps with no
internal time course. In the present model there are two natural
choices for the time at which weight changes occur: asynchronously
(instantaneously, whenever a presynaptic or postsynaptic spike
occurs), or synchronously (once per sweep of the repeated external
input, updating all weights simultaneously). We adopt the latter
strategy, updating weights at $x=0$ for each $t=nT, n \in
\mathbb{Z}$. The value of $w_i$ in the period beginning at $(0,t)$ is
then independent of $x$, and will be denoted $w_i(t)$. For synchronous
updating to be a reasonable approximation, we must assume that weight
changes per cycle are small relative to the weights themselves (slow
learning rate). Changes in weights due to different spikes or spike
pairs are assumed to add linearly.

In biological systems, synaptic weights have bounded magnitude and do
not change sign. Since the present paper is focused solely on the
dynamics near equilibria, we impose no boundary conditions in the
model. The results still apply to the biological case provided the
weight equilibria are in the region enclosed by biological bounds.

We assume homogeneous parameters: the scalar $\alpha$ and the
functions $\E$, $\L$ are the same for all presynaptic neurons, and the
times $x_i$ are regularly spaced, $x_i=i\delta$, $i=0,1,\ldots, N-1$ for
some $\delta>0$, $N=T/\delta \gg 1$.

For simplicity in the derivation of the weight dynamics, it will be
convenient to assume that $\E(s),\L(s)$ are zero or negligible for
$\abs{s}>\tau_E,\tau_L$ respectively, with $\tau_E,\tau_L \ll T$. We
will also require the learning rate to be slow: $T \ll \tau_w$, where
$\tau_w$ is the time-scale on which weights undergo significant
relative change. For the existence of approximate negative image
states we will need the spacing of presynaptic spike times much
smaller than the widths of $\E$ and $\L$: $\delta \ll
\tau_E,\tau_L$. These time-scale assumptions can be summarized as
\begin{equation}
\label{time-scales}
\delta \ll (\tau_E,\tau_L) \ll T \ll \tau_w.  \notag
\end{equation}   
Typical values for the mormyrid ELL are: $\delta < 1\text{ms}$
\cite{Bell92a}, [C.C. Bell, private communication], $\tau_E \sim
20\text{ms}$ \cite{Bell97a}, $\tau_L \sim 40\text{ms}$ \cite{Bell97a},
$T \sim 80\text{ms}$ [C.C. Bell, private communication], $\tau_w \sim
10^2T$ \cite{Bell97a}.

\section{Weight Dynamics}
\begin{figure}
\begin{center}
\newcommand{\sfw}{\figwidth}
\subfigure[]{\includegraphics[width=\sfw]{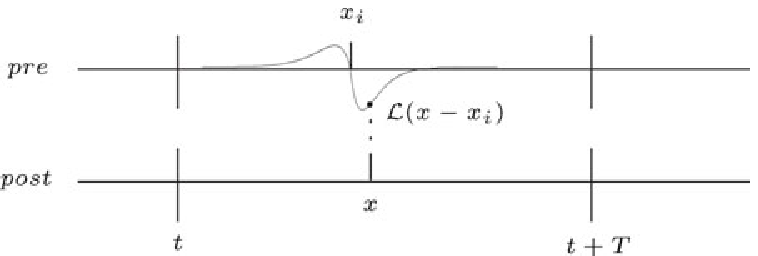}}
\subfigure[]{\includegraphics[width=\sfw]{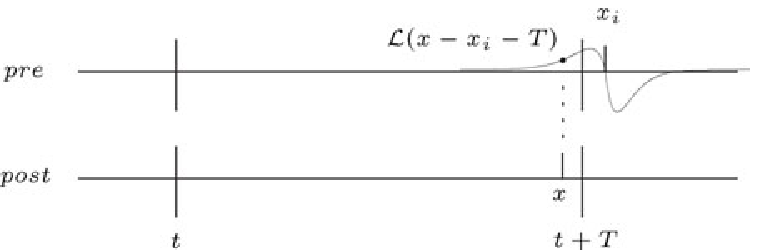}}
\caption{Changes in weight due to pairing of presynaptic and
  postsynaptic spikes. (a) Pairing of a postsynaptic spike at time
  $(x,t)$ and presynaptic spike by neuron $i$ at time $(x_i,t)$ causes
  a change $\L(x-x_i)$ in weight $w_i$. (b) For $x$ within $\tau_L$ of
  a period edge, we must include pairing with presynaptic spikes in
  the neighboring period. Pairing of a postsynaptic spike at time
  $(x,t)$ and presynaptic spike by neuron $i$ at time $(x_i,t+T)$
  cause a change $\L(x-x_i-T)$ in weight $w_i$.}
\label{figL}
\end{center}
\end{figure}

To obtain the mean weight dynamics, we compute the mean value of
$w_i(t+T)-w_i(t)$. The nonassociative change in $w_i(t)$ due to the
single presynaptic spike at $(x_i,t)$ is $\alpha$. For the associative
change due to presynaptic and postsynaptic spike pairs, consider the
effect of a single postsynaptic spike at $(x,t)$. The pairing of this
spike with the presynaptic spike at $(x_i,t)$ causes a change
$\L(x-x_i)$ in $w_i$. To properly handle edge effects, we also include
the pairing with presynaptic spikes at $(x_i,t-T)$ and $(x_i,t+T)$,
for a total change of
\begin{equation}
\label{threeL}
\L(x-x_i-T) + \L(x-x_i) + \L(x-x_i+T).
\end{equation}  
For typical biological applications, where $\tau_L \ll T$, at most one
of the above terms is non-negligible, but all must be included to
handle cases where $x-x_i$ is within $\tau_L$ of $T$ or $-T$
(Fig. \ref{figL}).
In addition, $\tau_L \ll T$ allows us to approximate
Eq. \eqref{threeL} by
\begin{equation}
\label{singlepost}
\sum_{-\infty}^{\infty} \L(x-x_i-nT)=\Lo(x-x_i),
\end{equation}
where $\Lo(s)=\sum_{-\infty}^{\infty} \L(s-nT)$ is the periodization
of $\L$ with period $T$.

\begin{figure}
\begin{center}
\newcommand{\sfw}{\figwidth}
\subfigure[]{\includegraphics[width=\sfw]{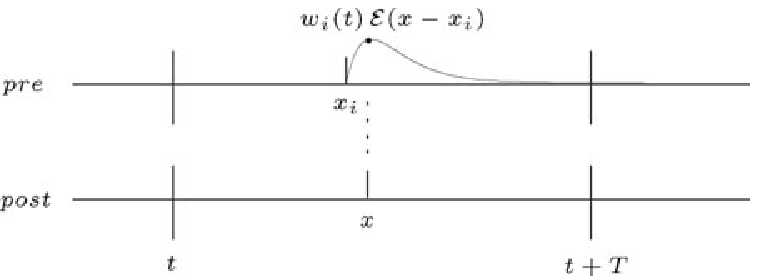}}
\subfigure[]{\includegraphics[width=\sfw]{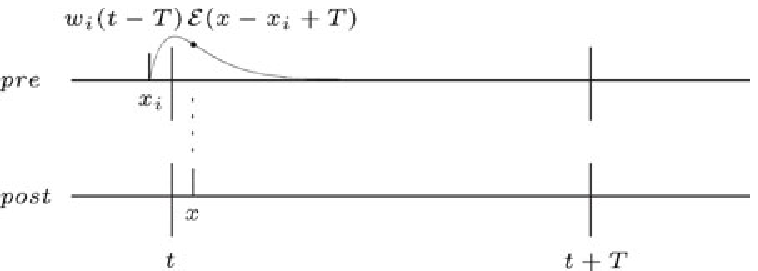}}
\caption{Postsynaptic potential due to presynaptic spikes. (a)
  Potential at time $(x,t)$ due to presynaptic spike by neuron $i$ at
  time $(x_i,t)$ is $w_i(t)\E(x-x_i)$. (b) For $x$ within $\tau_E$ of
  $0$, we must include the potential due to presynaptic spikes in the
  preceding period. The potential at time $(x,t)$ due to the
  presynaptic spike by neuron $i$ at time $(x_i,t-T)$ is
  {$w_i(t-T)\E(x-x_i+T)$.}}
\label{figE}
\end{center}
\end{figure}

Quantity \eqref{singlepost} is the change in $w_i(t)$ due to a single
postsynaptic spike at $(x,t)$. Postsynaptic spikes between $t$ and
$t+T$ occur at a mean rate density $f(U(x,t))$; hence the mean total
change due to all postsynaptic spikes between $t$ and $t+T$ is
\begin{equation}
\label{ }
\int_0^T dx\, f(U(x,t))\Lo(x-x_i).  \notag
\end{equation}
The mean total change in $w_i(t)$ due to both nonassociative and
associative learning is therefore
\begin{equation}
\label{meandeltaw}
\langle\triangle w_i(t)\rangle = \alpha + \int_0^T dx\,
f(U(x,t))\Lo(x-x_i). 
\end{equation}
We now compute the postsynaptic potential $U(x,t)$. The contribution
to $U(x,t)$ due to the presynaptic spike by neuron $i$ at $(x_i,t-nT)$
is $w_i(t+nT)\E(x-x_i+nT)$. For $\tau_E \ll T$ this quantity is
non-negligible for at most one value of $n$, either $n=0$ (current
period) or $n=-1$ (previous period). But to properly handle edge
effects (Fig. \ref{figE}) we include both, for a total contribution of
\begin{equation}
\label{twoE}
w_i(t-T)\E(x-x_i-T) + w_i(t)\E(x-x_i).
\end{equation}
We assume that the learning rate is sufficiently slow that we may
approximate quantity \eqref{twoE} by
\begin{equation}
\label{twoEonew}
w_i(t)(\E(x-x_i-T) + \E(x-x_i)).
\end{equation}
Finally, $\tau_E \ll T$ allows us to approximate quantity
\eqref{twoEonew} by
\begin{equation}
\label{singlepre}
w_i(t) \sum_{-\infty}^{\infty} \E(x-x_i-nT)=w_i(t) \Eo(x-x_i),
\end{equation}
where $\Eo(s)=\sum_{-\infty}^{\infty} \E(s-nT)$ is the periodization
of $\E$ with period $T$.

Quantity \eqref{singlepre} is the contribution to $U(x,t)$ from neuron
$i$. The total postsynaptic potential is the summed contribution from
all presynaptic neurons, plus the repeated external input:
\begin{equation}
\label{Uxt}
 U(x,t) = \phi(x)+\sum_{j=1}^N w_j(t)\Eo(x-x_j) 
\end{equation} 
Equations Eq. \eqref{meandeltaw} and Eq. \eqref{Uxt} define the mean
weight dynamics. The common periodicity of the functions $\Eo$, $\Lo$
and $\phi$ is an important feature, allowing the systematic use of
Fourier techniques.

\section{The Negative Image}

\begin{figure}
\begin{center}
\includegraphics[width=\figwidth]{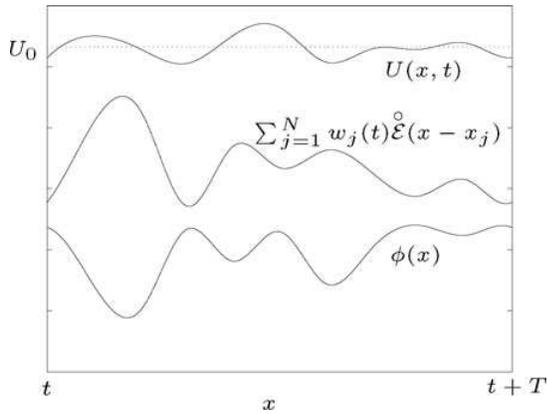}
\caption{An approximate negative image. If the postsynaptic potential
  $U(x,t)=\phi(x)+\sum_{j=1}^N w_j(t) \Eo(x-x_j)$ is approximately
  some constant $U_0$, then the potential $\sum_{j=1}^N w_j(t)
  \Eo(x-x_j)$ due to presynaptic spikes alone is approximately
  $U_0-\phi(x)$. }
\label{fignegim}
\end{center}
\end{figure}

A set of weights $\{w_i\}$ for which the total postsynaptic potential
$U(x,t)$ is approximately constant in $x$ will be referred to as an
{\it approximate negative image} state. For such a state the contribution to the
postsynaptic potential due to the presynaptic cells alone is, up to an
additive constant $U_0$, an approximate negative image
(Fig. \ref{fignegim}) of the external input $\phi$:
\begin{equation}
\label{approxnegimage}
\sum_{j=1}^N w_j \Eo(x-x_j)\simeq U_0-\phi(x).
\end{equation}
In the following, we first show that approximate negative image states
exist provided a certain condition holds on the Fourier coefficients
of the postsynaptic potential function $\Eo$ and the repeated external
input $\phi$, and provided the presynaptic spike time-spacing $\delta$
is sufficiently small. We then show that for a particular value of
$U_0$ (depending on $\alpha$, $\Lo$, and $f$) there exists an
approximate negative image state which is also an equilibrium (fixed
point) for the weight dynamics.

For generic $\Eo$ and $\phi$, Eq. \eqref{approxnegimage} cannot be
made an exact equality for all $x$, because that would require solving
infinitely many independent linear equations (one for each $x$) in
only finitely many unknowns (the N weights $\{w_j\}$). But if we
replace the discrete set of weights $w_j$ by a continuum weight
density $\W$, then the analog of Eq. \eqref{approxnegimage} can, under
certain conditions, be made exact for all $x$. Given such a density,
we then recover the biological case of discrete weights $\{w_j\}$ for
which Eq. \eqref{approxnegimage} is approximately true by defining the
set $\{w_j\}$ to be a discrete approximation to $\W$.

Let $\W(y)$ be a weight density, with $\W(y) dy$ being the total
weight for presynaptic spikes occurring between $y$ and $y+dy$, for $y
\in [0,T)$. The continuum analog of Eq. \eqref{approxnegimage}, with
  exact equality for all $x$, is
\begin{equation}
\label{contnegimage}
\int_0^T dy \W(y) \Eo(x-y) = U_0 - \phi(x).
\end{equation}
To solve this equation for $\W$ we take the Fourier decomposition. Let
$W_n=(1/T)\int_0^T dy\, e^{ik_ny}W(y)$ for $k_n=2\pi n/T$, $n\in
\mathbb{Z}$ be the Fourier coefficients for $\W$, and let $E_n$,
$\phi_n$ be the coefficients for $\Eo$ and $\phi$. Then
Eq. \eqref{contnegimage} becomes
\begin{eqnarray*}
U_0 & - & \sum_{n=-\infty}^{\infty} \phi_n e^{-ik_nx} \\ & = &
\int_0^T dy (\sum_{n=-\infty}^{\infty} W_n
e^{-ik_ny})(\sum_{m=-\infty}^{\infty} E_m e^{-ik_m(x-y)}) \\ & = &
\sum_{n=-\infty}^{\infty}\sum_{m=-\infty}^{\infty} W_nE_m e^{-ik_mx}
\int_0^T dy e^{i(k_m-k_n)y} \\ & = & T \sum_{n=-\infty}^{\infty}
W_nE_ne^{-ik_nx}.
\end{eqnarray*}
Hence $\W$ satisfies Eq. \eqref{contnegimage} if and only if
\begin{eqnarray}
W_0 & = & \frac{U_0-\phi_0}{TE_0}, \notag\\ W_n & = &
\frac{-\phi_n}{TE_n}, \quad n\neq 0.  \label{Wn}
\end{eqnarray}
Given such a $\W$, we construct approximate negative image states with
discrete weights as follows. Define $g(x)$ to be the deviation from a
negative image:
\begin{equation}
\label{discnegimage}
 g(x) = \phi(x) - U_0 - \sum_{j=1}^N w_j \Eo(x-x_j).
\end{equation}
Then $\{w_j\}$ is an approximate negative image state if $g(x)$ is
small relative to $U_0 - \phi(x)$, for all $x$. Consider the set of
weights defined by
\begin{equation}
\label{ }
w_j = \delta \,\,\W(x_j), \notag
\end{equation}
where $\delta$ is the spacing of the $x_j$. These weights can be
thought of as a discrete approximation to the weight density
$\W(y)$. Substituting into Eq. \eqref{discnegimage} and using
Eq. \eqref{contnegimage} gives
\begin{eqnarray*}
g(x) & = & \sum_{j=1}^N \delta \,\, \W(x_j) \Eo(x-x_j) - \int_0^T dy
\W(y) \Eo(x-y).
\end{eqnarray*}
This is the difference between a Riemann sum and the integral it
approximates. The error theorem for Riemann sums then gives an upper
bound for $g$:
\begin{equation}
\label{wigglebound}
\abs{g(x)} \leq \delta \frac{T}{2} \max_y
\abs{\frac{d}{dy}[\W(y)\Eo(x-y)]}.
\end{equation}
Hence, for $\abs{g(x)}$ to be small, we need $\W(y)\Eo(x-y)$ to be
differentiable in $y$, hence $\W(y)$ to be differentiable in $y$. A
theorem of Fourier series \cite{champeney} says that $\W(y)$ is
differentiable if $\sum_{n=-\infty}^{\infty} \abs{nW_n} < \infty$. By
Eq. \eqref{Wn} this places a constraint on the Fourier coefficients of
$\Eo$ and $\phi$:
\begin{equation}
\label{Wdiff}
\sum_{n=-\infty}^{\infty} \abs{\frac{n\phi_n}{E_n}} < \infty .
\end{equation}
This inequality requires $\phi_n$ to go to zero as $n\to\pm\infty$
more rapidly than $E_n/n^2$. In particular, the high frequency (large
$\abs{n}$) spectral content of $\phi$ must be less than the high
frequency content of $\Eo$. Intuitively, in order for the convolution
of $\Eo$ with a smooth weight density $\W$ to be able to ``match'' the
high frequency components of $-\phi$, the high frequency content of
$\phi$ cannot be too large.

If Eq. \eqref{Wdiff} is satisfied, and $\delta$ is sufficiently small,
then from Eq. \eqref{wigglebound} the deviation $g(x)$ from an exact
negative image is small, hence approximate negative image states
exist.

We now show that for a particular $U_0$ there exists an approximate
negative image state that is an equilibrium for the weight
dynamics. From Eq. \eqref{meandeltaw}, a weight state $\{w_j\}$ is an
equilibrium if $U(x)=\phi(x)+\sum_{j=1}^N w_j \Eo(x-x_j)$ satisfies
\begin{equation}
\label{eqUx}
\alpha + \int_0^T dx\, f(U(x)) \Lo(x-x_i)= 0 \quad \text{for all}
\,\,i.
\end{equation}
This is a system of $N$ equations in the $N$ unknowns $\{w_j\}$, but
they are nonlinear equations for nonlinear $f$. In general such
equations need not have solutions, but for approximate negative image
states the nonlinearity is in some sense ``small'', and this will
allow us to show that solutions exist provided $\delta$ is
sufficiently small.

For an approximate negative image state we have $U(x)=U_0+g(x)$ with
$g(x)\ll U_0$, and we wish this $U(x)$ to satisfy
Eq. \eqref{eqUx}. First define $U_0$ so that Eq. \eqref{eqUx} would be
satisfied if $g(x)$ were identically zero:
\begin{equation}
\label{eqU0}
\alpha + \int_0^T dx\, f(U_0) \Lo(x-x_i)= 0 \quad \text{for all}
\,\,i.
\end{equation}
This requires
\begin{eqnarray}
f(U_0) & = & \frac{-\alpha}{\int_0^T dx\, \Lo(x-x_i)} \quad \text{for
  all} \, i \notag \\ & = & \frac{-\alpha}{\int_0^T dx\, \Lo(x)},
\end{eqnarray}
where the independence of $i$ follows from the periodicity of
$\Lo$. Hence our desired $U_0$ exists and is given by
\begin{equation}
\label{U0}
U_0=f^{-1}\Bigl(\frac{-\alpha}{\int_0^T dx\, \Lo(x)}\Bigr),
\end{equation}
provided $\alpha$, $\Lo$ and $f$ satisfy
\begin{equation}
\label{alphaoverL}
\min_u f(u) < \frac{-\alpha}{\int_0^T dx\, \Lo(x)} < \max_u f(u).
\end{equation}
From Eq. \eqref{eqU0}, $U(x)=U_0+g(x)$ satisfies Eq. \eqref{eqUx} if
and only if
\begin{equation}
\label{eqUxminusU0}  
\int_0^T dx\, [f(U_0+g(x))-f(U_0)] \Lo(x-x_i) = 0 \quad \text{for all}
\,\,i.
\end{equation}
For brevity let $h(x)=f(U_0+g(x))-f(U_0)$ and
$L_i(x)=\Lo(x-x_i)$. Then Eq. \eqref{eqUxminusU0} can be written as
\begin{equation}
\label{hLi}
\langle h,L_i \rangle = 0 \quad \text{for all} \,\, i,
\end{equation}
where $\langle\cdot,\cdot\rangle$ is the inner product defined by
\begin{equation}
\label{innerprod}
\langle f_1,f_2 \rangle=\int_0^T dx\, f_1(x)f_2(x),
\end{equation}
for $f_1$, $f_2$ in the space $X$ of smooth functions on the interval
$[0,T]$.

Let $H$ be the set of functions $h$ corresponding to all possible
values of the weights $\{w_j\}$:
\begin{eqnarray*}
\label{ }
H=\{h: h(x)=f(\phi(x)-\sum_{j=1}^N w_j \Eo(x-x_j))-f(U_0),\\ \, w_j\in
\mathbb{R},\, j=1,\ldots,N\},
\end{eqnarray*} 
where we have used $U_0+g(x)=\phi(x)-\sum_{j=1}^N w_j \Eo(x-x_j)$ from
Eq. \eqref{discnegimage}. Let $S$ be the subspace of $X$ consisting of
all linear combinations of the $\{L_i\}$, and $S^{\bot }$ be the  (infinite dimensional) subspace
of $X$ orthogonal to $S$ (in the inner product defined by
Eq. \eqref{innerprod}). Then there exists an $h$ satisfying
Eq. \eqref{hLi} if and only if $H$ and $S^{\bot }$ have nonempty
intersection:
\begin{equation}
\label{HS}
H \cap S^{\bot } \ne \emptyset.
\end{equation}
We claim that condition \eqref{HS} holds if $\delta$ is sufficiently
small. If $\delta$ is small, then the bound \eqref{wigglebound}
implies that $g(x)$ is small for all $x$. In that case
$h(x)=f(U_0+g(x))-f(U_0)$ is approximately its linearization in $g$,
which we denote by $h_0(x)$:
\begin{eqnarray*}
\label{ }
h(x) \simeq h_0(x) & = & f'(U_0)g(x) \\ & = &
f'(U_0)(\phi(x)-U_0+\sum_{j=1}^N w_j \Eo(x-x_j)).
\end{eqnarray*} 
Let $H_0$ be the set of such $h_0$ corresponding to all possible
values of the weights $\{w_j\}$:
\begin{eqnarray*}
\label{ }
H_0=\{h_0: h_0(x)=f'(U_0)(\phi(x)-U_0+\sum_{j=1}^N w_j \Eo(x-x_j)),\\
\, w_j\in \mathbb{R},\, j=1,\ldots,N\},
\end{eqnarray*} 
Then the condition that $H_0$ have nonempty intersection with $S^{\bot
}$,
\begin{equation}
\label{H0S}
H_0 \cap S^{\bot } \ne \emptyset,
\end{equation}
is equivalent to existence of $h_0 \in H_0$ such that $\langle h_0,L_i
\rangle=0$ for all $i$. This is equivalent to the linearization of the
system \eqref{eqUxminusU0}:
\begin{eqnarray*}
\label{lineareqUxminusU0}
\int_0^T dx\, f'(U_0)(\phi(x)-U_0+\sum_{j=1}^N w_j \Eo(x-x_j))
\Lo(x-x_i) = 0 \\ \quad \text{for all} \,\,i,
\end{eqnarray*}
which can be rewritten as
\begin{equation}
\label{ }
\sum_{j=1}^N Q_{ij} w_j = \gamma,
\end{equation}
where $\gamma=\int_0^T dx\, f'(U_0)(\phi(x)-U_0)$ and
\begin{equation}
\label{Q}
Q_{ij}=f'(U_0)\int_0^T dx\, \Eo(x-x_j) \Lo(x-x_i).
\end{equation} 
This is a system of $N$ linear inhomogeneous equations in the $N$
unknowns $\{w_j\}$, which has a solution provided the coefficient
matrix $Q$ is invertible. The eigenvalues of $Q$ will be calculated in
the following section, and for generic $\E$ and $\L$ all eigenvalues
are nonzero. Hence $Q$ is generically invertible, so that condition
\eqref{H0S} holds. Furthermore, the intersection of $H_0$ with
$S^{\bot }$ is generically transversal (not tangent).

\begin{figure}
\begin{center}
\includegraphics[width=\figwidth]{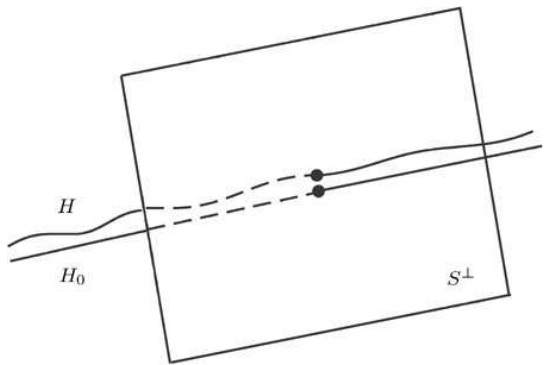}
\caption{Transversal intersection theorem. If $H_0$ has transversal
  intersection with $S^{\bot }$ and $H$ is sufficiently close to
  $H_0$, then $H$ intersects $S^{\bot }$.}
\label{figtransversal}
\end{center}
\end{figure}

Now as $\delta\to 0$, $H \to H_0$ (in the metric induced by the inner
product \eqref{innerprod}). By the openness of transversal
intersection (infinite dimensional version, \cite{AbrahamRobbin}), any sufficiently small perturbation
of $H_0$ also intersects $S^{\bot }$. Hence for $\delta$ sufficiently
small, $H$ intersects $S^{\bot }$ (Fig. \ref{figtransversal}), hence
$h$ satisfying Eq. \eqref{hLi} exists. The corresponding weight state
$\{w_j\}$ is an approximate negative image equilibrium.

\section{Stability Criterion}

We now derive a necessary and sufficient condition for the mean
stability of approximate negative image equilibria, by examining the
linearized weight dynamics around such states. Let $\{\hat{w}_j\}$ be
an approximate negative image equilibrium satisfying
Eq. \eqref{discnegimage} with
\begin{equation}
\label{Uxwhat}
U(x)=\phi(x)+\sum_{j=1}^N \hat{w}_j\Eo(x-x_j)= U_0+\hat{g}(x)
\end{equation}
Solving for $\phi(x)$ in Eq. \eqref{Uxwhat} and substituting into
\eqref{Uxt} yields
\begin{eqnarray*}
\label{ }
U(x,t) &=& U_0 + \hat{g}(x) + \sum_{j=1}^N
v_j(t)\Eo(x-x_j), \notag
\end{eqnarray*}
where $v_j(t)=w_j(t)-\hat{w}_j$ is the deviation of weight $j$ from
its equilibrium value, and $\hat{g}(x)$ is the deviation from a
negative image in the equilibrium state $\{\hat{w}_j\}$. To first
order in $v_j$ we then have
\begin{equation}
\label{fUxtv}
f(U(x,t)) \simeq f(U_0+\hat{g}(x)) + f'(U_0+\hat{g}(x))\sum_{j=1}^N
v_j(t)\Eo(x-x_j).
\end{equation}
Substituting Eq. \eqref{fUxtv} into Eq. \eqref{meandeltaw} and using
$\triangle w_i(t)= \triangle v_i(t)$ yields
\begin{eqnarray*}
\label{deltameanvfirst}
\langle \triangle v_i(t) \rangle = \alpha + \int_0^T dx\, f(U_0+
\hat{g}(x))\Lo(x-x_i) \notag \\ + f'(U_0+ \hat{g}(x)) \sum_{j=1}^N
v_j(t)\Eo(x-x_j))\Lo(x-x_i).
\end{eqnarray*}
From the equilibrium condition Eq. \eqref{eqUx} and
Eq. \eqref{Uxwhat}, the term in Eq. \eqref{deltameanvfirst} of zeroth
order in $v_j$ vanishes. Hence
\begin{equation}
\label{ }
\langle \triangle v_i(t) \rangle \simeq \sum_{j=1}^N P_{ij} v_j(t),
\notag
\end{equation}
where
\begin{equation}
\label{ }
P_{ij} = \int_0^T dx\, f'(U_0+
\hat{g}(x))\Eo(x-x_j))\Lo(x-x_i). \notag
\end{equation}
Now assume $\delta$ is sufficiently small that
\begin{equation}
\label{ }
\hat{g}(x) \ll \frac{f'(U_0)}{f''(U_0)} \quad \text{for all }
x. \notag
\end{equation}
Then $f'(U_0+ \hat{g}(x)) \simeq f'(U_0)$ for all $x$, so that $P_{ij}
\simeq Q_{ij}$, where $Q$ is the matrix defined in Eq. \eqref{Q}, and
we obtain
\begin{equation}
\label{deltameanvalmost}
\langle \triangle v_i(t) \rangle \simeq \sum_{j=1}^N Q_{ij}
v_j(t). \notag
\end{equation}
Taking the mean on both sides and using $\langle \triangle v_i(t)
\rangle = \triangle \langle v_i(t) \rangle$ yields
\begin{equation}
\label{deltameanv}
\triangle \langle v_i(t) \rangle \simeq \sum_{j=1}^N Q_{ij} \langle
v_i(t) \rangle.
\end{equation}
Eq. \eqref{deltameanv} gives the linearized dynamics for $\langle
v_i(t) \rangle$ near $\langle v_i(t) \rangle=0$, hence for $\langle
w_i(t) \rangle$ near the approximate negative image equilibrium
$\{\hat{w}_j\}$.

The system Eq. \eqref{deltameanv} is stable if and only if all
eigenvalues of $Q+I$ have norm less than 1. Due to periodicity of
$\Eo,\Lo$ and regular spacing of the $\{x_i\}$, the matrix $Q$ has the
property that each of its rows equals the row above it shifted one
entry to the right (and wrapped around at the edges). Such matrices
are called {\it circulant} \protect\cite{davis} and their eigenvectors
and eigenvalues are easily found, as follows. Let $u$ be the vector
with components $u_i=e^{ikx_i}$; then
\begin{eqnarray*}
\label{ }
(Qu)_i &=& \sum_{j=1}^N Q_{ij}u_j \\ &=& \sum_{j=1}^N
e^{ikx_j}\int_0^T dx\, \Eo(x-x_j) \Lo(x-x_i) ,
\end{eqnarray*}
so that
\begin{equation}
\label{Quoveru}
\frac{(Qu)_i}{u_i}=\sum_{j=1}^N e^{ik(x_j-x_i)}\int_0^T dx\,
\Eo(x-x_j) \Lo(x-x_i).
\end{equation}
By periodicity of $\Eo$ and $\Lo$, the integral in the above
expression is a function of $x_j-x_i$ modulo $T$. The factor
$e^{ik(x_j-x_i)}$ is also a function of $x_j-x_i$ modulo $T$ provided
$e^{ikT}=1$. Now if the $\{x_i\}$ are regularly spaced, the sum over
$j$ of a function of $x_j-x_i$ modulo $T$ is independent of $i$. In
that case Eq. \eqref{Quoveru} would imply $(Qu)_i/u_i$ independent of
$i$, hence $u$ is an eigenvector of $Q$, with eigenvalue the right
hand side of Eq. \eqref{Quoveru}. We get a complete set of such
eigenvectors by taking $N$ values of $k$ such that $e^{ikT}=1$ and the
functions $e^{ikx_i}$ are independent functions of $i$. Here we choose
\begin{equation}
\label{ }
k_n=\frac{2\pi n}{T}, \quad n=0,1,...,N-1.  \notag
\end{equation} 
The corresponding eigevalues of $Q$ are
\begin{eqnarray*}
 \lambda_n & = & \sum_{j=1}^N e^{ik_n(x_j-x_i)}\int_0^T dx\,
 \Eo(x-x_j)\Lo(x-x_i).
\end{eqnarray*}   
Letting $z_j=x_j-x_i$, making the change of variables $y=x_j-x$ and
using periodicity of $\Eo$, $\Lo$ gives
 \begin{eqnarray*}
 \lambda_n & = & \sum_{j=1}^N e^{ik_nz_j}\int_0^T dy
 \Eo(-y)\Lo(z_j-y)\\ & = & \sum_{j=1}^N e^{ik_nx_j} \Lo *_T
 \widetilde{\Eo}\, (x_j),
\end{eqnarray*} 
where $*_T$ is convolution on the interval $[0,T]$, $\widetilde{\,\,}$
is horizontal reflection ($\widetilde{\Eo}(y)=\Eo(-y))$, and since
$\{z_j\}=\{x_j\}$ we have replaced $z_j$ by $x_j$ in the sum.

The stability condition is $\abs{1+\lambda_n}<1$ for all $n$:
\begin{gather}
\text{{\it Stability:}} \notag\\
\label{discdiscstab} \abs{1+\sum_{j=1}^N e^{ik_nx_j} \Lo *_T \widetilde{\Eo}\, (x_j)}<1,
\\ \quad k_n=\frac{2\pi n}{T}, \, n=0,1,...,N-1. \notag
\end{gather} 
In the biological setting two limiting regimes are of special
interest: slow learning ($\Lo$ small) and dense spacing ($\delta$
small).

If $\Lo$ is small, then so are the eigenvalues of $Q$. If
$\lambda_n=a_n+ib_n$ with $a_n,b_n$ real, we have
\begin{equation}
\label{ }
\abs{1+\lambda_n}^2=(1+a_n)^2-b_n^2=1+2a_n +(a_n^2-b_n^2). \notag
\end{equation}
If $a_n$ and $b_n$ are sufficiently small, this quantity is less than
$1$ if and only if $a_n<0$. Hence for $\Lo$ sufficiently small, all
eigenvalues of $Q+I$ have norm less than $1$ if and only if all
eigenvalues of $Q$ have negative real part\footnote{The slow learning
  limit can thus be thought of as the continuous time (continuous $t$)
  limit. All eigenvalues of $Q$ having negative real part is
  equivalent to stability of the system $d\langle v\rangle/dt=Q\langle
  v\rangle$ and hence of $T\,d\langle v\rangle/dt=Q\langle v\rangle$,
  which is the continuous time version of \protect
  Eq. \eqref{deltameanv}.}. The stability condition then becomes $Re\,
\lambda_n <0$ for all $n$. Hence in the slow learning limit
Eq. \eqref{discdiscstab} becomes
\begin{gather}
\text{{\it Slow learning:}} \notag\\ Re \sum_{j=1}^N e^{ik_nx_j} \Lo
*_T \widetilde{\Eo}\, (x_j) < 0, \quad
n=0,1,...,N-1.  \label{slowstab}
\end{gather}

The dense spacing limit ($\delta \to 0$) is the continuum limit in
$x_i$. The discrete weight density $w_i/T$ is replaced by a continuum
weight density $W(x)$, sums over $x_j$ are replaced by integrals over
$x$, and $N\to \infty$. This yields
\begin{eqnarray*}
\label{ }
\lambda_n & = &\int_0^T dx\, e^{ik_nx}\Lo *_T \widetilde{\Eo}\, (x)
\quad n=0,1,...
\end{eqnarray*}
Hence $\lambda_n$ is just the $n^{th}$ Fourier coefficient of $\Lo *_T
\widetilde{\Eo}$. The Fourier convolution theorem then gives
\begin{equation}
\label{ }
\lambda_n = \Lro_n \widetilde{\Ero}_n = \Lro_n \overline{\Ero_n},
\notag
\end{equation}
where $\Ero_n, \widetilde{\Ero}_n, \Lro_n$ are the $n^{th}$ Fourier
coefficients of $\Ero, \widetilde{\Ero}, \Lro$ respectively, and
$\overline{z}$ is the complex conjugate of $z$. Substituting into the
stability condition $\abs{1+\lambda_n}<1$ for all $n$ gives the dense
spacing limit of Eq. \eqref{discdiscstab}:
\begin{gather}
\text{{\it Dense spacing:}} \notag\\
\label{densestab} \abs{1+\Lro_n \overline{\Ero_n}}<1, \quad n=0,1,...  
\end{gather}
Finally, with both slow learning and dense spacing the stability
condition becomes
\begin{gather}
\text{{\it Slow learning, dense spacing:}} \notag\\ Re [\Lro_n
  \overline{\Ero_n}] < 0, n=0,1,... \label{slowdensestab}
\end{gather}

A further simplification follows in the long period limit,
$\tau_E,\tau_L \ll T$. Holding $\tau_E,\tau_L$ constant and taking $T
\to \infty$, the Fourier series of $\Eo,\Lo$ in
Eq. \eqref{slowdensestab} approach Fourier transforms of $\E,\L$. The
stability condition then becomes
\begin{gather}
\text{{\it Slow learning, dense spacing, long period:}} \notag\\ Re
     [\F{\L}(k)\overline{\F{\E}}(k)] < 0, \quad k\in (-\infty,\infty).
\label{slowdenselongstab}
\end{gather}
For the calculation of examples we will work in the slow learning,
dense spacing, long period limit, which is the limit of primary
biological interest in the mormyrid ELL.

\section{General Remarks}

\textbf{The Roles of Nonassociative and Associative Learning. } Both
nonassociative and associative learning ($\alpha$ and $\L$,
respectively) play a role in whether approximate negative image
equilibria exist, via Eq. \eqref{alphaoverL}. They are also both
involved in determining the location of such equilibria, via
Eq. \eqref{eqUx}. The interpretation of Eq. \eqref{eqUx} is that at
equilibrium, the mean change due to nonassociative learning ($\alpha$)
must be precisely opposite to the mean change due to associative
learning (the $\L$ term). If the postsynaptic spike rate density $f$
is bounded, this places a relative magnitude constraint on $\alpha$
and $\L$, namely Eq. \eqref{alphaoverL}. If this constraint is
violated then the mean changes due to associative and nonassociative
learning are unable to balance one another, and no negative image
equilibrium is possible.

By contrast, only associative learning plays a role in the stability
of approximate negative image equilibria, via
Eq. \eqref{discdiscstab}. The irrelevance of nonassociative learning
for stability has an intuitive interpretation: near an approximate
negative image equilibrium, the mean nonassociative change is
cancelled by the mean associative change due the constant postsynaptic
potential $U_0$ around which $U(x,t)$ fluctuates. Only the deviations
of $U(x,t)$ from $U_0$ cause a net change in the weights, and these
changes are purely associative (due to postsynaptic spikes generated
by $U(x,t)$). Alternatively, nonassociative learning can be analogized
to a constant externally applied force in a physical system. Such a
force changes the location of equilibria, but has no effect on the
dynamics around equilibria.

\textbf{The Role of the Repeated Input. } For a given postsynaptic
potential response $\E$, the repeated input $\phi$ plays a role in the
existence of approximate negative image states via Eq. \eqref{Wdiff}:
$\phi$ cannot have too much high frequency content relative to $\E$
for such states to exist.

Assuming $\phi$ is such that approximate negative image states exist,
it then plays the important role of determining the weight
configurations in such states, an in particular in 
approximate negative image equilibria, via Eq. \eqref{approxnegimage}.

But $\phi$ plays no role in the stability of the resulting negative
image equilibrium. This is intuitively reasonable, since in
approximate negative image states $\phi$ is ``nulled out'' by the
summed postsynaptic potentials due to time-locked presynaptic spikes.

\textbf{The Role of Noise. } The functional form of the mean
postsynaptic spike rate $f$ affects the existence and location of the
negative image equilbrium via Eq. \eqref{alphaoverL} and
Eq. \eqref{eqUx}.

But provided $f$ is strictly increasing (so that $f'$ is
positive), $f$ has no effect on the stability of the
equilibrium. Hence the classification of learning rules as (mean)
stable or unstable is, except for this mild monotonicity requirement,
insensitive to the fine structure of the noise. This is a post hoc
justification for not modeling the noise in more detail.

\textbf{Canonically Stable Learning Rule: $\L=-\E$. } In the dense
spacing and slow learning limit, suppose $\L=-\E$. The stability
condition Eq. \eqref{slowdensestab} is then
\begin{equation}
\label{}
\abs{\Ero_n}^2 > 0 \qquad \text{for all} \,\, n,
\end{equation}
or in other words, $\Ero_n \ne 0$ for all $n$.  Since this is true for
generic $\E$, the learning rule $\L=-\E$ is generically stable.

\textbf{Area Sign Condition. } In the dense spacing and slow
learning limit, consider $n=0$ in Eq. \eqref{slowdensestab}. Since
$\Lro_0$ and $\overline{\Ero_0}=\Ero_0$ are just the areas under the
functions $\Lo$ and $\Eo$, Eq. \eqref{slowdensestab} says that for
stability, these areas must be opposite in sign. If they are the
same sign, then the negative image is unstable. In particular, if $\E$
and $\L$ are both nonnegative, the negative image is unstable. Hence,
if $\E$ is an excitatory PSP and $\L$ is any pure potentiating
learning rule, the negative image is unstable. Similarly, inhibitory
PSPs with purely depressing learning rules are unstable.

\textbf{Symmetric and Antisymmetric Learning Rules. } In the dense
spacing, slow learning, long period limit, there is a nonempty,
positive measure set of postsynaptic response functions for which
purely symmetric or purely antisymmetric learning rules are
generically unstable. This follows from the fact that the Fourier
transforms of symmetric and antisymmetric functions are pure real and
pure imaginary, respectively.

\begin{figure}
\begin{center}
\includegraphics[width=\figwidth]{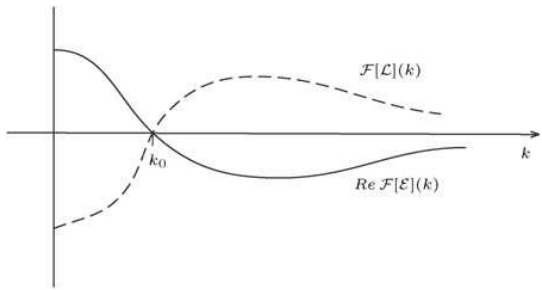}
\caption{If $Re[\F{\E}(k)]$ changes sign at $k_0$, then for the
  product $\F{\L}(k) Re [\F{\E}(k)]$ to be negative around $k_0$ we
  must have $\F{\L}(k)$ also change sign at $k_0$, in the opposite
  sense to $Re[\F{\E}](k)$. }
\label{figk0}
\end{center}
\end{figure}

Suppose the real part of the Fourier transform of $\E$ has a zero:
\begin{equation}
\label{refek0}
Re[\F{\E}(k_0)]=0, \quad \text{for some } k_0.
\end{equation}
Then, generically, $Re[\F{\E}(k)]$ changes sign at $k_0$. Suppose $\L$
is symmetric, so that $\F{\L}(k)$ is pure real. Then
\begin{equation}
\label{stabquant}
Re[\F{\L}(k)\overline{\F{\E}}(k)]=\F{\L}(k)\, Re[\F{\E}(k)]. \notag
\end{equation}
Since $Re[\F{\E}(k)]$ changes sign at $k_0$, for the stability
condition Eq. \eqref{slowdenselongstab} to be satisfied for $k$ near
$k_0$, we must have $\F{\L}(k)$ change sign at $k_0$, in the opposite
sense to $Re[\F{\E}(k)]$; see Fig. \ref{figk0}. But this forces
$\F{\L}(k_0)=0$, which is untrue for generic symmetric $\L$. Hence
generic symmetric learning rules are unstable for postsynaptic
response functions satisfying Eq. \eqref{refek0}.

Similarly, if the imaginary part of the Fourier transform of $\E$ has
a zero:
\begin{equation}
\label{imfek0}
Im[\F{\E}(k_0)]=0, \quad \text{for some } k_0.
\end{equation}
then generic antisymmetric learning rules are unstable.

Pure antisymmetric learning rules have another difficulty: since they
satisfy $\int_0^T dx\, \L(x) = 0$, near a negative image equilibrium
the mean weight change per cycle due to an antisymmetric $\L$ is zero,
to first order in $g$. The total mean weight change per cycle is
therefore approximately $\alpha$. Hence negative image equilibria for
pure antisymmetric learning rules are only possible if $\alpha=0$ (no
nonassociative learning).

\textbf{Cooperative Stability. } It follows from
Eq. \eqref{discdiscstab} that the sum of stable learning rules is
stable; but it is also clear that given a generic $\E$, there exist
pairs of learning rules $\L_1$ and $\L_2$, each individually unstable,
for which the sum $\L_1 + \L_2$ is stable. This is most easily seen by
direct computation in the slow learning, dense spacing, long period
limit, via Eq. \eqref{slowdenselongstab} (see the examples calculated
below).

\textbf{Duality Principle. } Interchanging $\L$ and $\E$ in
Eq. \eqref{Q} transforms $Q_{ij}$ to $Q_{ji}$, hence $Q$ to $Q^T$,
hence $Q+I$ to $(Q+I)^T$. The eigenvalues of a matrix are unchanged by
transposition. The stability condition, that all eigenvalues of $Q+I$
have norm less than $1$, is thus invariant under interchange of $\L$
and $\E$. In other words, a PSP $\E$ and learning rule $\L$ are a
stable pair if and only if the PSP $\L$ and learning rule $\E$ are a
stable pair.

This has potential biological relevance if the functional forms of
PSPs and associative learning rules overlap. The single-lobe
exponential and alpha function learning rules treated in the examples,
below, are also plausible PSPs, hence duality applies.

\textbf{Inversion Principle. } Replacing $\E$ by $-\E$ and $\L$ by
$-\L$ in Eq. \eqref{Q} leaves $Q_{ij}$ invariant, hence $Q+I$
invariant. The stability condition is therefore invariant under
inversion of both $\E$ and $\L$. In other words, a PSP $\E$ and
learning rule $\L$ are a stable pair if and only if the PSP $-\E$ and
learning rule $-\L$ are a stable pair.

In particular, the stable learning rules for an inhibitory PSP are
just minus the stable learning rules for the corresponding excitatory
PSP. Plasticity at inhibitory synapses was explored in
\cite{Roberts00d}, and preliminary experimental evidence given in
\cite{Han99a}.

\textbf{Independence of Normalization. } In the slow learning and
dense spacing limit, the stability conditions
Eq. \eqref{slowdensestab} or Eq. \eqref{slowdenselongstab} are
invariant under multiplication of $\L$ or $\E$ by positive
constants. Hence, provided the magnitudes of $\L$ or $\E$ are not so
large that the slow learning assumption is violated, stability does
not depend on those magnitudes. In particular, in working with
specific examples it is not necessary to give $\L$ or $\E$ any overall
normalization.

\begin{figure}
\begin{center}
\newcommand{\sfw}{\figwidth}
\subfigure[]{\includegraphics[width=\sfw]{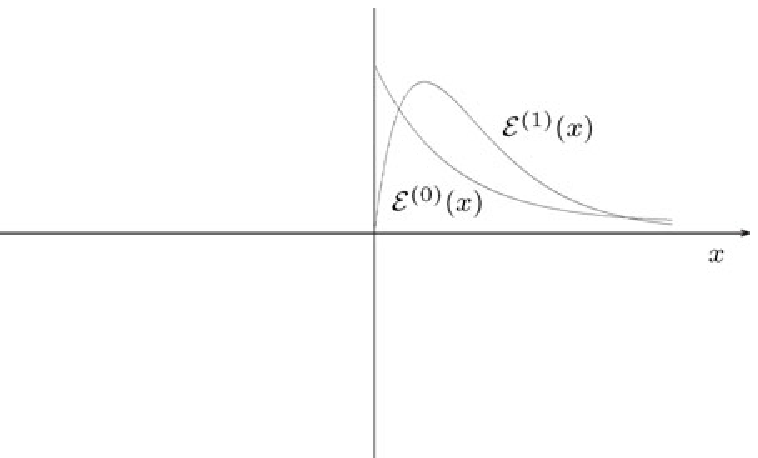}}
\subfigure[]{\includegraphics[width=\sfw]{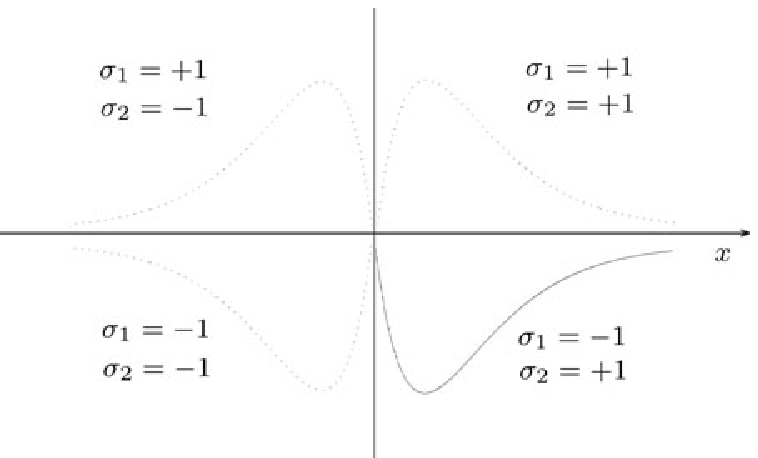}}
\subfigure[]{\includegraphics[width=\sfw]{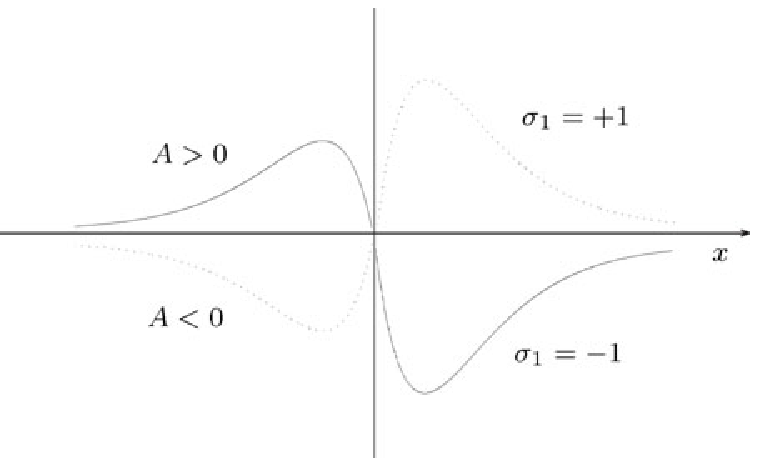}}
\caption{PSPs and learning rules used in the examples. (a) The PSP
  $\E^{(p_E)}(x)$ is exponential for $p_E=0$ and an alpha function for
  $p_E=1$. (b) One-lobe learning rules of alpha function form,
  $\L_{I}^{(1)}(x)$, for the four possible combinations of $\sigma_1$
  (potentiating or depressing) and $\sigma_2$ (pre-before-post or
  post-before-pre).  (c) Two-lobe learning rules of alpha function
  form, $\L_{II}^{(1)}(x)$, for the four possible combinations of
  $\sigma_1$ (pre-before-post lobe potentiating or depressing) and the
  sign of $A$ (post-before-pre lobe potentiating or depressing). The
  area of the pre-before-post lobe is normalized to $\pm 1$, and the
  area of the post-before-pre lobe is $A$. }
\label{figELonetwo}
\end{center}
\end{figure}

\section{Examples}

Working in the slow learning, dense spacing, long period limit, we now
compute explicit criteria for stability when $\E$ and $\L$ have
functional forms commonly used in the spike-timing dependent
plasticity literature. The PSP $\E$ will be assumed excitatory and
causal, and of exponential or alpha function form. The learning rule
$\L$ will consist of one or two ``lobes'': a ``pre-before-post'' lobe
(presynaptic spike before postsynaptic spike) and/or a
``post-before-pre'' lobe (postsynaptic spike before presynaptic
spike). Each lobe will be of exponential or alpha function form, and
either potentiating (positive) or depressing (negative). Such $\E$ and
$\L$ can be written as follows:
\begin{align*}
\E^{(p_E)}(x) &= x^{p_E} e^{-x/\tau_E} H(x) \\ \L_{I}^{(p_L)}(x) &=
\sigma_1 \, (\sigma_2x)^{p_L} e^{-\sigma_2 x/\tau_L} H(\sigma_2x)
&\text{(one-lobe)} \\ \L_{II}^{(p_L)}(x) &=
\frac{\sigma_1}{\tau_{L_1}^{p_L+1}} \, x^{p_L} e^{-x/\tau_{L_1}}
H(x)\\ & \quad + \frac{A}{\tau_{L_2}^{p_L+1}} \,(-x)^{p_L}
e^{x/\tau_{L_2}} H(-x) &\text{(two-lobe)}
\end{align*}
where $H$ is the Heaviside function
\begin{equation}
\label{ }
H(x)=\begin{cases} 1 & x\ge 0, \\ 0 & x<0.
\end{cases}    \notag
\end{equation}
The parameters (see Fig. \ref{figELonetwo}) are as follows:
$p_E,p_L=0$ for an exponential or $1$ for an alpha function;
$\tau_E$, $\tau_L$, $\tau_{L_1}$ and $\tau_{L_2}$ are positive time
constants; $\sigma_1=+1$ for a potentiating lobe or $-1$ for a
depressing lobe; $\sigma_2=+1$ for a pre-before-post lobe or $-1$ for
a post-before-pre lobe; and for the two-lobe $\L$, $A$ is the area of
the post-before-pre lobe, with the area of the pre-before-post lobe
normalized to $\pm 1$ . We impose no overall normalization on $\E$ or
$\L$, since this has no effect on stability.

We assume an excitatory PSP $\E$; to obtain the stable cases for the
inhibitory PSP $-\E$, simply replace $\L$ by $-\L$ in the stable cases
for $\E$ (i.e. replace $\sigma_1$ by $-\sigma_1$ and $A$ by $-A$).

For both the one-lobe and two-lobe $\L$, there are four possible
combinations of $p_E$ and $p_L$: exponential or alpha function PSP
with exponential or alpha function learning rule. We will refer to
these four cases as \textbf{ee}, \textbf{ea}, \textbf{ae}, and
\textbf{aa}, with the first letter in the pair indicating that the PSP
is exponential or alpha function, and the second letter referring to
the learning rule.

The Fourier transforms of these $\E$ and $\L$ are rational functions
in $k$, and the stability condition will reduce to the requirement
that a certain polynomial in $k^2$, whose coefficients are themselves
polynomials in the parameters of $\E$ and $\L$, be negative for all
$k$. Since the algebra in all cases is essentially the same, differing
only in the size and coefficients of the resulting polynomial, we
present only one case (\textbf{aa}, one-lobe) in full detail and for
all other cases simply list the end results.

\subsection{Alpha Function PSP, One-Lobe Alpha Function Learning Rule}
\begin{figure}
\begin{center}
\includegraphics[width=\figwidth]{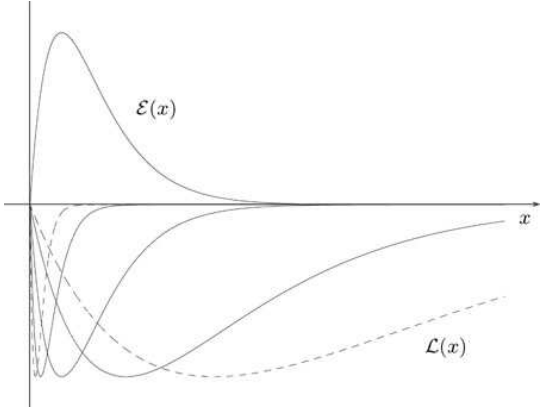}
\caption{Range of stable one-lobe $\L$ for given $\E$, in case
  \textbf{aa}. The learning rule must be depressive and
  pre-before-post, with $3-2\sqrt{2}<
  \tau_L/\tau_E<3+2\sqrt{2}$. Stable examples are drawn with
  solid lines; endpoints of the stable interval are drawn with dashed
  lines.}
\label{figLonerange}
\end{center}
\end{figure}

For
\begin{eqnarray*}
\E(x)& = &xe^{-x/\tau_E}H(x)\\ \L(x)& = &\sigma_1 \sigma_2 xe^{-\sigma_2
  x/\tau_L}H(\sigma_2 x)
\end{eqnarray*}
we have
\begin{eqnarray*}
\F{\E}(k)& = &\frac{\tau_E^2}{(1-ikE)^2}\\ \F{\L}(k)& = & \frac{\sigma_1
  \tau_L^2}{(1-\sigma_2 ik\tau_L)^2}
\end{eqnarray*}
leading to
\begin{eqnarray*}
&&Re\Big\{\F{\L}(k)\overline{\F{\E}}(k)\Big\} \\ &=& C \,
Re[\sigma_1(1+i\sigma_2k\tau_L)^2(1-ik\tau_E)^2]\\ &=& C
\sigma_1[\sigma_2^2\tau_L^2\tau_E^2k^4 +
  (4\sigma_2\tau_L\tau_E-\sigma_2^2\tau_L^2-\tau_E^2)k^2+1]
\end{eqnarray*}
where $C=\tau_L^2
\tau_E^2/[(1+\sigma_2^2\tau_L^2k^2)^2(1+\tau_E^2k^2)^2]$. Since $C>0$,
the stability condition is then
\begin{eqnarray}
\label{aa1cond}
\sigma_1[\sigma_2^2r^2k^4 + (4\sigma_2r-\sigma_2^2r^2-1)k^2+1] < 0
\notag \\ \quad \text{for all} \,\, k
\end{eqnarray}
where $r=\tau_L/\tau_E$. The expression on the left is a quadratic in
$k^2$. The condition is impossible for $\sigma_1=+1$ (it fails at
$k=0$) but for $\sigma_1=-1$ more work is required. The quadratic
$ax^2+bx+c$ is negative for all $x\ge 0$ if and only if $a<0$ and
($b<0$ or $b^2-4ac<0$). Applying this condition to Eq. \eqref{aa1cond}
with $\sigma_1=-1$ yields
\begin{eqnarray}
 r^2-4\sigma_2r+1& < &0\\ \text{or} \quad
 (r-\sigma_2)^2(r^2-6\sigma_2r+1) & < & 0
\end{eqnarray}
For $\sigma_2=-1$ these give $-2-\sqrt{3}<r<-2+\sqrt{3}$ or
$-3-2\sqrt{2}<r<-3+2\sqrt{2}$, both of which are impossible because
$r>0$. For $\sigma_2=+1$ we get $2-\sqrt{3}<r<2+\sqrt{3}$ or
$3-2\sqrt{2}<r<3+2\sqrt{2}$; the former is contained in the latter,
giving stability if and only if
\begin{eqnarray}
\label{finalaa1cond }
\sigma_1=-1, \sigma_2=+1, \notag \\ 3-2\sqrt{2}<
\frac{\tau_L}{\tau_E}<3+2\sqrt{2}.
\end{eqnarray} 
The only stable case is depressive and pre-before-post, with
$\tau_L/\tau_E$ constrained to lie in a finite interval
(Fig. \ref{figLonerange}).  Note that this interval contains
$\tau_L/\tau_E=1$, where $\L=-\E$ (the canonically stable learning
rule).

Duality is also applicable here. Interchanging $\E$ and $\L$ in this example is
equivalent to interchanging $\tau_E$ and $\tau_L$ and multiplying both
$\E$ and $\L$ by $-1$. The multiplications offset and we are left with
$r$ replaced by $1/r$. It follows that if the interval of stability
for the pair $(\E,\L)$ is $s_1<r<s_2$, the corresponding interval for
the pair $(\L,\E)$ is $s_1<1/r<s_2$. But by duality these intervals
must coincide; hence we must have $s_1=1/s_2$. This is indeed the case
for $s_1=3-2\sqrt{2}$ and $s_2=3+2\sqrt{2}$.

The instability of the $\sigma_1=+1$ case for any $\tau_E$ and
$\tau_L$, by the failure of the stability condition at $k=0$, is just
the Area Sign Condition.

\begin{figure*}
\begin{center}
\newcommand{\sfw}{\figwidth}
\subfigure{\includegraphics[width=\sfw]{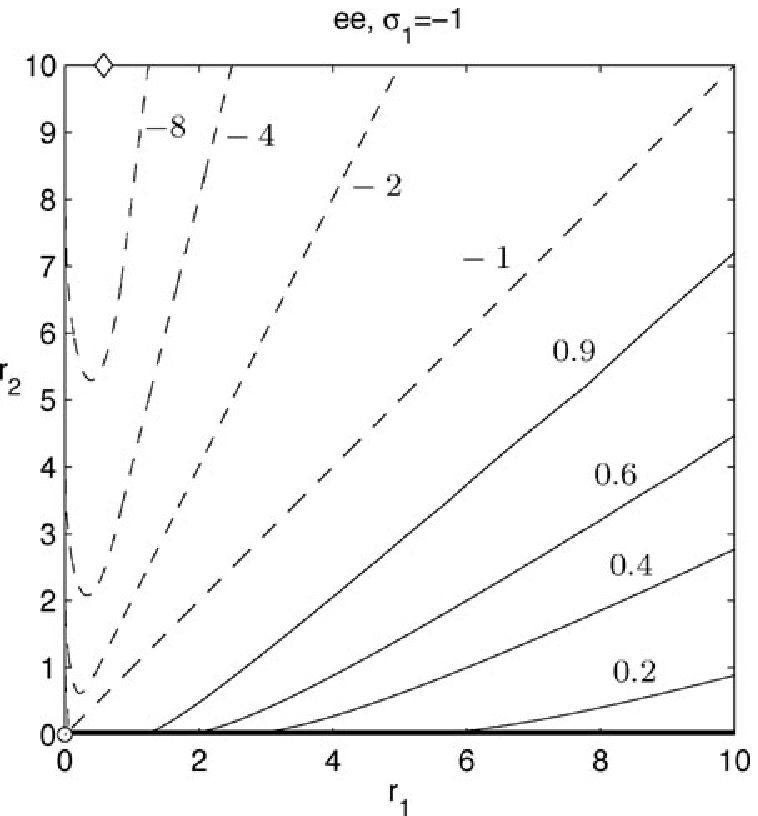}}
\subfigure{\includegraphics[width=\sfw]{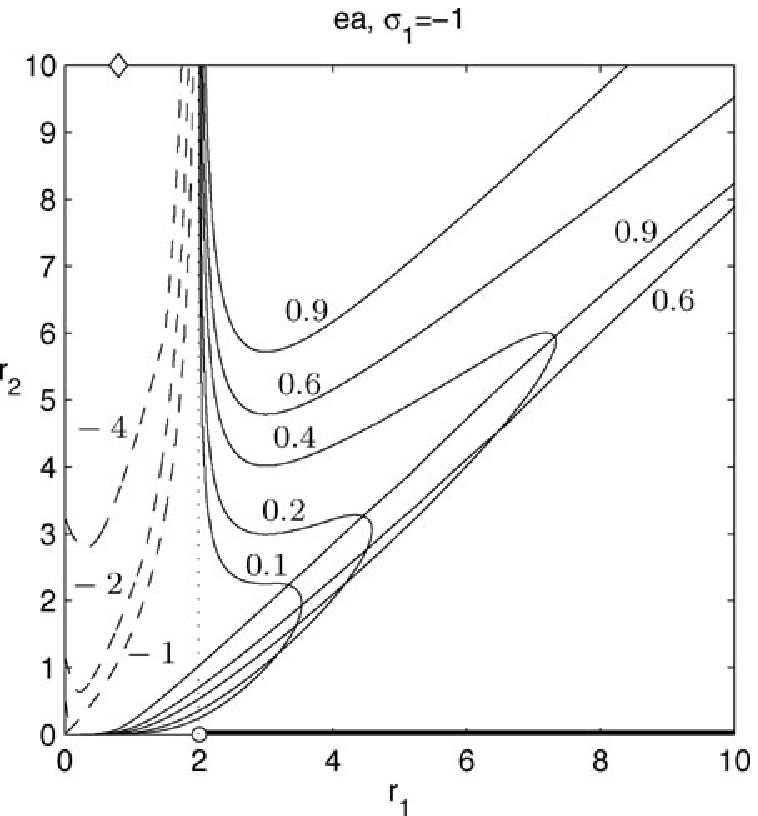}}
\subfigure{\includegraphics[width=\sfw]{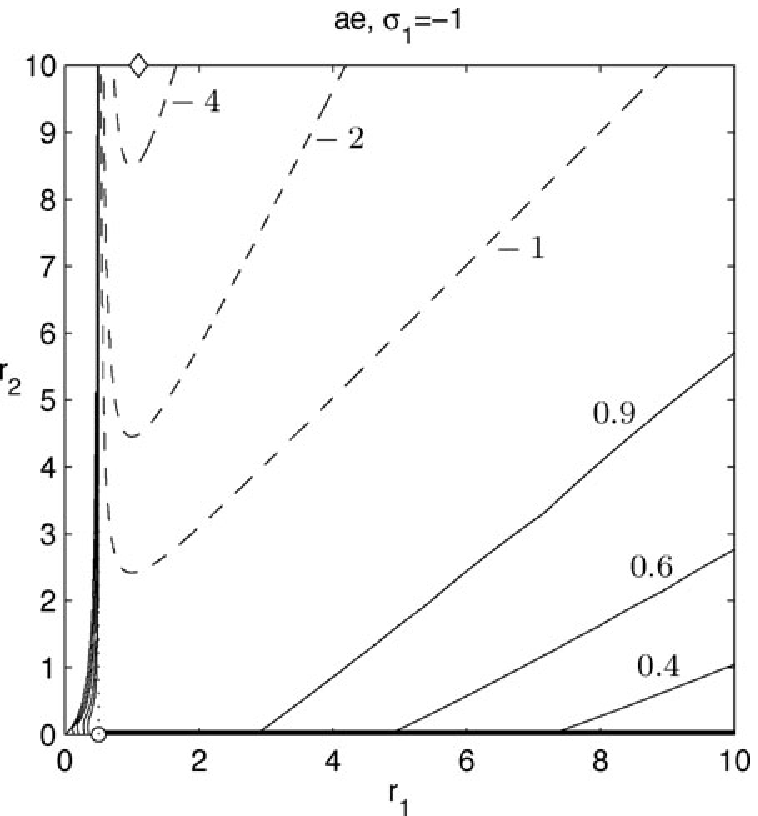}}
\subfigure{\includegraphics[width=\sfw]{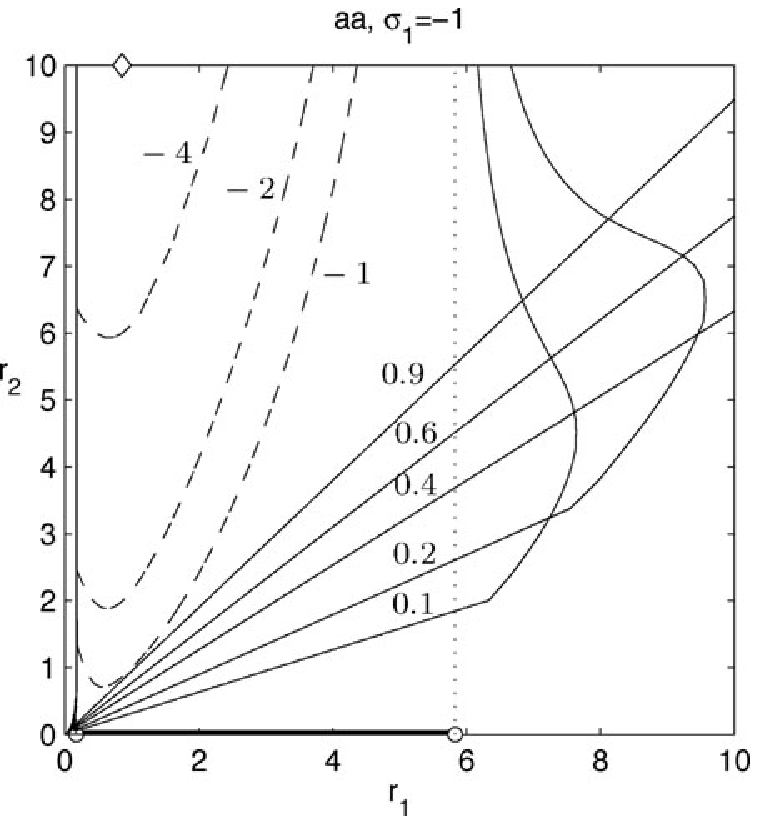}}
\caption{Boundary curves of the stable region in $(r_1,r_2)$ for
  various values of $A$, for two-lobe $\L$ with $\sigma_1=-1$. Curves
  are labelled by $A$. Curves with $A>0$ are drawn with solid lines,
  curves with $A<0$ with dashed lines. In all cases the region of
  stability is on the side of the curve containing the diamond
  ($\diamond$) in the upper left corner of the plot. The interval of
  stability for the corresponding one-lobe learning rules is the
  portion of the $r_1$-axis in bold.  }
\label{fig2dplots}
\end{center}
\end{figure*}

\subsection{Summary of Results}
For the one-lobe learning rules the stable parameter ranges are all
easily calculated:
\begin{description}
  \item[\textbf{ee}:] \quad $\sigma_1=-1,\sigma_2=+1$: \quad all
    $\tau_L/\tau_E$
  \item[\textbf{ea}:] \quad $\sigma_1=-1,\sigma_2=+1$: \quad
    $\tau_L/\tau_E<2$
  \item[\textbf{ae}:] \quad $\sigma_1=-1,\sigma_2=+1$: \quad
    $\tau_L/\tau_E>1/2$
  \item[\textbf{aa}:] \quad $\sigma_1=-1,\sigma_2=+1$:\quad
    $2-\sqrt{3}<\tau_L/\tau_E<2+\sqrt{3}.$
\end{description}
Note that in all four cases we get instability, for all $\tau_L$ and
$\tau_E$, if $\L$ is not depressive and pre-before-post. For $\L$
depressive and pre-before-post, all four cases have some range of
$\tau_L/\tau_E$ in which $\L$ is stable. The extent of that range
depends critically on the precise functional form of $\E$ and $\L$;
but for $1/2 < \tau_L/\tau_E <2$ we have stability independent of the
functional form of $\E$ and $\L$.

For the two-lobe learning rules the polynomial arising out of the
stability condition has coefficients depending on $\sigma_1$ and on
three continuous parameters: $r_1$, $r_2$ and $A$, where
$r_1=\tau_{L_1}/\tau_E$, $r_2=\tau_{L_2}/\tau_E$. The polynomials are
given in the Appendix.

In all four cases, $\sigma_1=+1$ is always unstable. For
$\sigma_1=-1$, the boundaries of the stable region in $(r_1,r_2)$ for
various values of $A$ are plotted numerically in
Fig. \ref{fig2dplots}.

For one-lobe learning rules we found that only depressive and
pre-before-post permits stability. For two-lobe learning rules, the
pre-before-post lobe must be depressive for stability, and the
post-before-pre lobe cannot have area $A$ greater than 1. This is just
the Area Sign Condition: the area of the pre-before-post lobe is -1,
and for stability when paired with an excitatory PSP $\E$ the total
area under the learning rule $\L$ must be negative.

For $A<1$, the effect of the post-before-pre lobe shows the following
general trends: in cases \textbf{ee} and \textbf{ae}, as the absolute
area $\abs{A}$ of the post-before-pre lobe increases, the stable region
in the relative time constants $r_1$ and $r_2$ tends to shrink. Hence
the post-before-pre lobe can be thought of as destabilizing in such
cases. In cases \textbf{ee} and \textbf{ae} the situation is less
clear. Increasingly negative $A$ (larger depressive post-before-pre)
is uniformly destabilizing, but increasingly positive $A$ (larger potentiating post-before-pre) appears to
be destabilizing for small $r_2$ but stabilizing for large $r_2$.

Cooperative stability, in which a two-lobe rule is stable while each
of its lobes individually would be unstable, occurs in cases
\textbf{ea}, \textbf{ae}, and \textbf{aa}: any point $(r_1,r_2)$ in a
stable region, with $r_1$ outside the interval in which the
corresponding one-lobe rule is stable is an example of cooperative stability.

Finally, the shape and extent of stable regions for two-lobe learning
rules, or the extent of stable intervals for one-lobe learning rules,
depend critically on whether $\E$ and $\L$ are exponential or alpha
function in form. This suggests that in order to infer even such
qualitative properties as stability or instability in a biological
context, the learning rule must be known with considerable precision.

However, for particular values of some parameters the
dependence on functional form may be such that useful conclusions can
still be drawn in the absence of such precision; for example, the stability in one-lobe, depressive
pre-before-post learning rules with $\tau_L/\tau_E \simeq 1$,
independent of whether $\E$ or $\L$ are exponential or alpha function
in form. This particular finding has direct relevance to the learning
rule observed experimentally in mormyrid ELL \cite{Bell97a}. The
experimental data is not precise enough to suggest a particular
functional form, but does indicate a one-lobe, depressive,
pre-before-post rule, with a width of the same order of magnitude as
the width of a PSP. Stability of such a rule is consistent with the
analytic results derived above.

\section{Appendix}
For completeness we provide below the polynomial conditions for
stability of the two-lobe learning rules treated in the examples.
\begin{description}
  \item[\textbf{ee}:] \quad $ak^4+bk^2 +c < 0 \quad \text{for all}
    \,\, k$\\ \quad $ a = \sigma_1r_1r_2^2-Ar_1^2r_2$\\ \quad $ b =
    \sigma_1(r_2^2+r_1) + A(r_1^2-r_2)$\\ \quad $ c = \sigma_1+A$

\item[\textbf{ea}:] \quad $ak^6+bk^4 +ck^2 +d < 0 \quad \text{for
    all} \,\, k$\\ 
\quad $ a = \sigma_1r_1r_2^3(2r_2-1)
    -Ar_1^3r_2(2r_1+1)$\\ \quad $ b =
    \sigma_1r_2(r_2^3-3r_1^2r_2+6r_1r_2+r_1)+Ar_1(r_1^3-3r_1r_2^2-6r_1r_2+r_2)$\\
    \quad $ c = \sigma_1(2r_2^2-r_1^2+2r_1)+A(2r_1^2-r_2^2-2r_2)$\\
    \quad $ d = \sigma_1+A$
 
  \item[\textbf{ae}:] \quad $ak^4+bk^2 +c < 0 \quad \text{for all}
    \,\, k$\\ \quad $ a = \sigma_1r_2^2(2r_1-1) -Ar_1^2(2r_2+1)$\\
    \quad $ b = \sigma_1(r_2^2+2r_1-1)+A(r_1^2-2r_2-1)$\\ \quad $ c =
    \sigma_1+A$

  \item[\textbf{aa}:] \quad $ak^8+bk^6 +ck^4 +dk^2+e< 0 \quad
    \text{for all} \,\, k$\\ \quad $ a = \sigma_1r_1^2r_2^4 +
    Ar_1^4r_2^2$\\ \quad $ b =
    \sigma_1(-r_2^4-2r_1^2r_2^3+4r_1r_2^4-r_1r_2^3+3r_1^2r_2^2+2r_1r_2^2)\\
    +Ar_1(-r_1^4+2r_1^3r_2^2-4r_1^4r_2-r_1^3r_2=2r_1^2r_2+3r_1^2r_2^2)$\\
    \quad $ c =
    \sigma_1(-3r_1r_2+(r_1^2+r_2^2)(r_2+1)^2+r_2^2(1+4r_1-4r_1r_2))\\
    +A(-3r_1r_2+(r_1^2+r_2^2)(r_1-1)^2+r_1^2(1-4r_2-4r_1r_2))$\\ \quad
    $d = \sigma_1(2r_2^2-r_1^2+4r_1-1)+A(2r_1^2-r_2^2-4r_2-1)$\\ \quad
    $ e = \sigma_1+A$
\end{description}

\begin{acknowledgments}
We would like to thank Dr. Gerhard Magnus, Dr. Nathaniel Sawtell, and
the members of Dr.\ Curtis Bell's lab for insightful discussions.
This material is based upon work supported by the National Science
Foundation under Grant No. IBN-0114558, and by the National Institute of Mental Health under Grant No. R01-MH60364.
\end{acknowledgments}

\bibliography{bigrefs}

\end{document}